\documentclass[preprint,12pt]{elsarticle}

\usepackage{amsmath}
\usepackage[english]{babel}
\usepackage{graphicx}

\usepackage{color}

\journal{ArXiv (article in Physica A)}

\begin{document}

\begin{frontmatter}

\title{Transition in fluctuation behaviour of normal liquids under high pressures}

\author{Eugene B. Postnikov}
\ead{postnicov@gmail.com}
\address{Theoretical Physics Department, Kursk State University, Radishcheva st., 33 Kursk 305000, Russia}

\author{Miros{\l}aw Chor{\c a}{\.z}ewski}
\ead{miroslaw.chorazewski@us.edu.pl}
\address{Institute of Chemistry, University of Silesia, Szkolna 9, 40-006 Katowice, Poland}

\begin{abstract}
We explore the behaviour of the inverse reduced density fluctuations and the isobaric expansion coefficient using $\alpha,\omega$-dibromoalkanes as an example. Two different states are revealed far from the critical point: the region of exponentially decaying fluctuations near the coexistence curve and the state with longer correlations under sufficiently high pressures. The crossing of the isotherms of the isobaric expansion coefficient occurs within the PVT range of the mentioned transition. We discuss the interplay of this crossing with the changes in molecular packing structure connected with the analysed function of the density, which represents inverse reduced volume fluctuations.

{\it \bf Highlights:}

\begin{itemize}
\item There are two regions with the different fluctuation behaviour

\item Fluctuation transition is determined by the random loose/close packing 

\item The crossing of expansivity isotherms coincides with this transition
\end{itemize}

\end{abstract}


\end{frontmatter}

\section{Introduction}

The isobaric coefficient of thermal expansion is a thermoelastic property defined as $\alpha_p\equiv-\rho^{-1}\left(\partial \rho/\partial T\right)_P$ and related to the thermal response of the system. This definition of the expansion coefficient refers both to a solid and a liquid phase. While the theory of the thermal expansion of solids is well developed and it is known that the thermal expansion in solid condensed systems is mainly caused by the anharmonicity of oscillations\cite{Ho}, in the case of liquids, the understanding of this thermodynamic function in terms of molecular behaviour is still a challenge. 

Extremely interesting is that the pressure dependence of the isotherms of thermal expansivity can exhibit anomalous behaviour in the case of simple liquids, i.e. isotherms present crossing in some characteristic pressure range, in which $\left(\partial \alpha_p/\partial T\right)_P=0$. Although the first observation of this phenomenon is dated to the work by  P.G. Bridgman \cite{Bridgman1913}, its comprehensive studies have been evaluated during the last decades only \cite{Randzio1994,Troncoso2009,Taravillo2003,Tellez2011,Chorazewski2010,Troncoso2011,Randzio2014}. 
The curve, at which the derivative with respect to temperature vanishes, divides the $p-T$ plane into two regions where $\left(\partial \alpha_p/\partial T\right)_P>0$  and $\left(\partial \alpha_p/\partial T\right)_P<0$. The latter corresponds to the anomalous volume expansion response to the increasing temperature.

Some single-phase equations of state, which can reproduce such a behaviour, as a rule postulate this crossing {\it a priory} and introduce several adjustments to this fact, as the modified repulsive contribution to the Carnahan-Starling model \cite{Randzio1995}, the shifted Lennard-Jones pair potentials\cite{Deiters1995} or the spinodal hypothesis \cite{Taravillo2003,Chauhan2013,Alba1985,Baonza1993}.

In our study, we consider the isothermal equation of state based on the continuation of the fluctuation law detected for a saturated liquid into the single-phase region.
This approach results in the crossing of the thermal expansivity isotherms without any additional assumptions. As well, we discuss its background from the point of view of liquid structure characteristics. 

\section{The crossing of expansivity isotherms and molecular packing}

The fluctuation model is based on the consideration of the inverse magnitude of reduced volume fluctuations. This quantity is determined as an inverse ratio of relative volume fluctuations in a condensed medium to their value for the hypothetical ideal gas at the same PVT thermodynamic conditions:
\begin{equation}
\nu=\left[\left.\frac{\left\langle (\Delta V)^2\right\rangle}{V}\right/\frac{\left\langle (\Delta V)^2_{ig}\right\rangle}{V_{ig}}\right]^{-1}=\frac{\mu_0}{RT}\left[\rho\beta_T\right]^{-1},
\label{nu}
\end{equation}
where $\mu_0$, $R$, $T$, $\rho$, $\beta_T$ are the molar mass, the gas constant, the temperature, the mass density and the isothermal compressibility correspondingly.

It has been observed \cite{Goncharov2013} that this parameter for simple real liquids and the model lattice fluid has the exponential dependence on the density 
\begin{equation}
\nu=\exp(\kappa\rho+b) 
\label{nuexp}
\end{equation}
along the coexistence curve well below the critical point curve . The coefficients $\kappa$, $b$ have a very weak dependence on the temperature and could be considered as constants with a reasonable accuracy \cite{Goncharov2013,Postnikov2014,Chorazewski2015}.

The assumption of the same functional dependence the single-phase region of liquids taken under elevated pressure allows for a simple integration of Eq.~(\ref{nu}) with the substituted (\ref{nuexp}). This procedure results in the two-parametric  Fluctuation-based Tait-like isothermal Equation of State (FT-EoS)
\begin{equation}
\rho=\rho_0+\frac{1}{\kappa}\log\left[\frac{\kappa\mu_0}{\nu(\rho_0)RT}(P-P_0)+1\right],
\label{rhop}
\end{equation}
which can be used for a prediction of the density along an isotherm \cite{Postnikov2014}. Its further differentiation with respect to the temperature provides the expression for the isobaric expansion coefficient \cite{Chorazewski2015}
\begin{equation}
\alpha_p=\alpha_p^0\frac{\rho_0}{\rho}e^{-\kappa(\rho-\rho_0)}+\frac{1-e^{-\kappa(\rho-\rho_0)}}{\kappa\rho T},
\label{alphc}
\end{equation}
Here in (\ref{rhop}) and (\ref{alphc}) the index $0$ marks the values taken at the referent state: the data taken along the coexistence curve or at the normal pressure well below the critical point. 

Now let us show that the properties of Eq.~{(\ref{alphc}) imply an existence of the crossing  $\left(\partial \alpha_p/\partial T\right)_P=0$ without any artificial assumptions. The derivative of $\alpha_p$ defined by Eq.~{(\ref{alphc}) with respect to the temperature at a constant pressure gives
\begin{eqnarray}
\left(\frac{\partial \alpha_p}{\partial T}\right)_P&=\alpha_p^2+
\frac{1}{\rho}\left\{-\frac{1}{\kappa T^2}+\left[\rho_0\left(\left(\frac{\partial \alpha_p^0}{\partial T}\right)_P-{\alpha_p^0}^2\right)+\right.\right.& \label{dalpha}\\
&\left.\left.
\frac{1}{\kappa T^2}+\kappa\left(\rho_0\alpha_p^0-\frac{1}{\kappa T}\right)(\rho\alpha_p-\rho_0\alpha_p^0)\right]
e^{-\kappa(\rho-\rho_0)}\right\}&\nonumber
\end{eqnarray}

For $\rho=\rho_0$ it reduces to $\left(\partial \alpha_p^0/\partial T\right)_P>0$. At the high-density asymptotics $\rho>>\rho_0$, the exponential factor multiplied by the square bracket in (\ref{dalpha}) tends to zero and the rest terms,
\begin{equation}
\left(\frac{\partial \alpha_p}{\partial T}\right)_P=\alpha_p^2-\frac{1}{\rho\kappa T^2}<0
\label{dminus}
\end{equation}
taking into account the characteristic orders of the included thermodynamic values for the studied organic liquids such as n-alkanes and their halogenated.

Rather, Eq.~(\ref{dminus}) provides an opportunity to check whether or not the crossing occurs based on the saturated data for $\alpha_p$ and $\rho$ of the compressed liquid calculated from Eq.~(\ref{rhop}) (or some other well-established isothermal approximation, say the classical Tait equation with known coefficients). 

However, Eq.~(\ref{dminus}) is a crude estimator, which indicates the fact of existence (or  nonexistence). The practical calculation of the proper density region requires an application of the full formula (\ref{dalpha}) with the substituted density determined by Eq.~(\ref{rhop}) under the given elevated pressure $P$. It should be also noted that Eq.~(\ref{dalpha}) is the point-wise one with respect to the temperature, i.e. it implies the calculation along a single isotherm if the referent $\alpha_P$ is known. 

The present model was tested using experimental data concerning the density, the speed of sound and the thermal expansivity of $\alpha,\omega$-dibromoalkanes (1,3-dibromopropane to 1,6-dibromohexane\cite{Chorazewski2015}, dibromomethane\cite{Chorazewski2014-2}, and the results of recent new measurements for 1,2-dibromoethane \cite{Chorazewski2015-2}, which complete this series), a polar non-associated liquid in which important electrostatic intermolecular interactions occur due to the permanent dipole moments of the molecules. In addition, these  compounds are characterized by the so-called intramolecular proximity effect\cite{Kehiaian1983}, according to which a change in the distance between two halogen atoms in the molecule causes a change in the molecular properties. 

Fig.~\ref{fgr:intersect} shows intervals, where the crossing occurs, for the series of $\alpha,\omega$-dibromoalkanes. They are quite close to the ones found experimentally \cite{Chorazewski2015,Chorazewski2014-2} although we use only the data measured at the normal conditions as referent ones. 

The results shown in Fig.~\ref{fgr:intersect} confirm also general behaviour observed primarily in a homologous series of alkanes\cite{Troncoso2011}, that the pressure range where all isotherms of thermal expansion coefficient intersect each other has a shift towards the lower-pressure region with increasing carbon atoms number in the alkane chain. 

\begin{figure}[t]
\centering
  \includegraphics[width=\columnwidth]{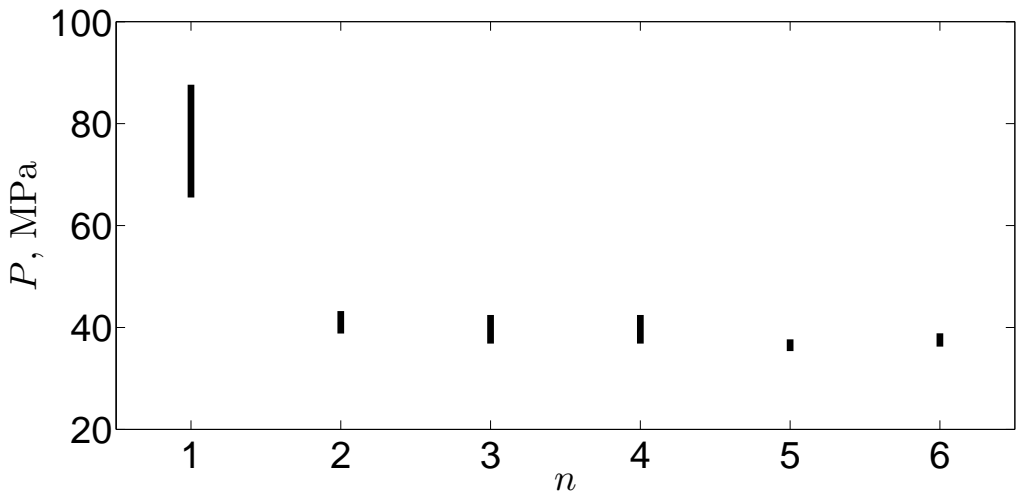}
  \caption{The calculated intervals of pressure, where crossing of the isobaric expansivities occurs for $\alpha,\omega$-dibromoalkanes as a function of the number of carbon atoms. 
	\label{fgr:intersect}}
\end{figure}

Now let us consider the validity of Eq.~(\ref{nuexp}) as a fit of the experimental data under elevated pressure in more details. Fig.~\ref{fgr:allnu} shows this exponential approximation in comparison with the inverse reduced fluctuation calculated from the experimental isothermal compressibilities, densities and temperatures.  

One can see that all data points corresponding to densities below the ones denoted by the vertical dotted lines fall onto straight lines, which represent the coexistence-based exponential fit in semi-logarithmic co-ordinates.
 Further, the inverse reduced fluctuations depend on both the temperature and the density. The densities corresponding to these vertical lines correspond to the pressures of the crossing of the isobaric expansivity isotherms\cite{Chorazewski2015} shown in Fig.~\ref{fgr:intersect}.

\begin{figure*}[t]
\centering
  \includegraphics[width=\textwidth]{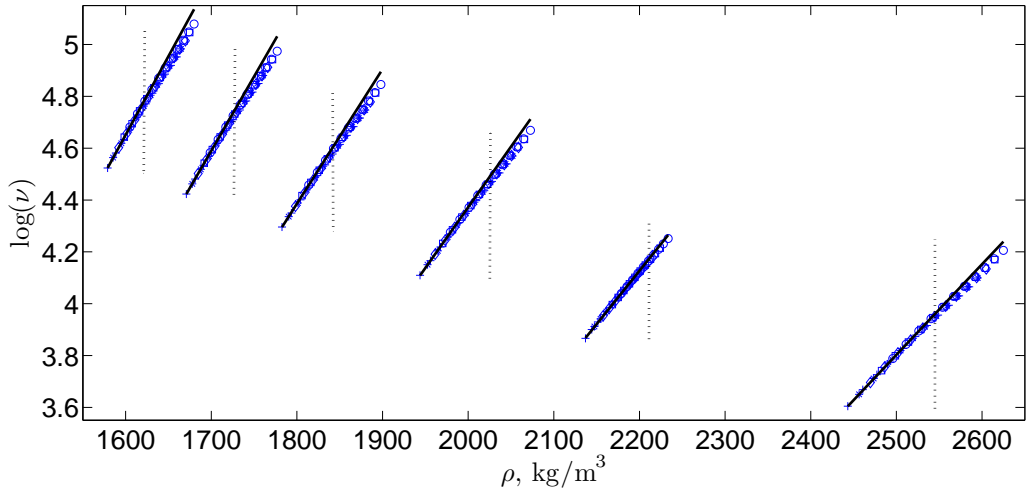}
  \caption{The inverse reduced fluctuations $\nu(\rho)$ functions of the density at elevated pressure for the series of $\alpha,\omega$-dibromoalkanes (from left to right: 1,6-dibromohexane, 1,5-dibromopentane, 1,4-dibromopropane, 1,3-dibromobutane, 1,2-dibromoethane, dibromomethane) at the temperatures 293.15~K (circles), 298.15~K (squares), 303.15~K (diamonds), 308.15~K (asterisks), 313.15~K (pluses). Vertical dotted lines denote the boundary of deviations from linear universality (the next point deviation exceeds an uncertainty of the linear fit), and solid lines are the trends calculated using saturated data.
	\label{fgr:allnu}}
\end{figure*}

Let us explore the transition between the exponential and non-exponential behaviours using dibromomethane as an example. 
It has been argued \cite{Goncharov2013} that the mentioned exponential dependence of the fluctuation parameter emerges in the saturated mean-field lattice fluid model due to geometric packing constraints. The available data on dibromomethane allow for discussing of packing properties for a real liquid. 

AIChE DIPPR 801 database provides $V_{vdW}= 2.268243\cdot10^{-4}\,\mathrm{m^3/kg}$ as the value of the van der Waals reduced volume for this substance. The packing density is determined as $\phi=\rho V_{vdW}$. Therefore, the density range presented in Fig.~\ref{fgr:numeth} corresponds to $\phi\in[0.554,\,0.595]$, which lies within the range of sphere packing from random loose packing (RLP) $\phi_{RLP}=0.555\pm0.005$ to random close packing (RCP) $\phi_{RCP}=0.635\pm0.005$ \cite{Onoda1990}. 

At the same time, the detailed study of hard spheres random packing \cite{Aste2008} revealed two additional structural changes at intermediate packing fractions $\phi_{i_1}\sim0.58$ and $\phi_{i_2}\sim0.6$. The characteristic interval of losing exponential universality denoted by the dotted and the dash-dotted lines in Fig.~\ref{fgr:numeth} corresponds to $\phi_T\in[0.572,\, 0.576]$, i.e. close to the first of the listed values. Since the acentric factor of dibromomethane is not very large (0.20945) and $\phi_T$ is far enough from RCP, it is admissible to refer to systems of spherical particles.   

Note that for the more realistic model system, the Weeks-Chandler-Andersen (WCA) fluid\cite{Heyes2006}, which takes into account inter-particle pair interactions with the truncated and shifted Lennard-Jones potential, it has been found that the rigidity percolation transition also occurs at $\phi\approx0.58$, the value coinciding with the glass transition of the hard-sphere liquid. 

\begin{figure}[h!]
\centering
  \includegraphics[width=\columnwidth]{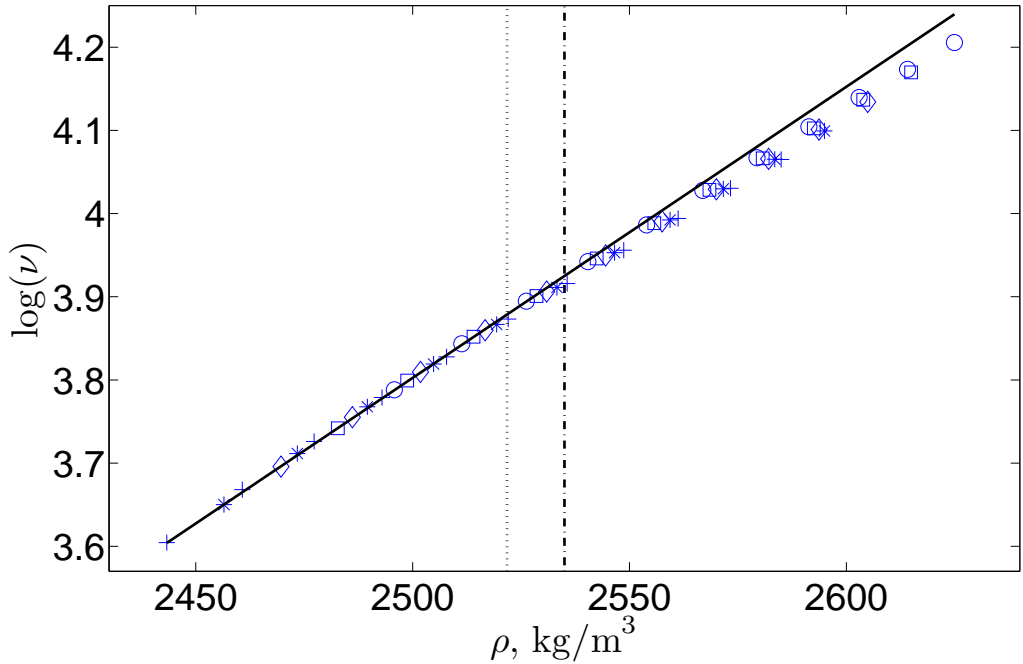}
  \caption{The inverse reduced fluctuations $\nu(\rho)$ in dibromomethane as a function of the density for various temperatures: 293.15~K (circles), 298.15~K (squares), 303.15~K (diamonds), 308.15~K (asterisks), 313.15~K (pluses). The solid line $\kappa=0.003499$, $b=-4.944$
corresponds to the function $\nu=\exp(\kappa\rho+b)$ fitting the coexistence curve data. The vertical lines bound the transition region (see the text).
		\label{fgr:numeth}}
\end{figure}

However, neither WCA fluid nor dibromomethane exhibits vitrification around $\phi\approx0.57-0.58$ due to a softeness of the interaction potential. The mentioned transition simply implies the proximity of  nearest-neighbour molecules.   As a result, the elastic properties of a fluid ($\nu\sim(\partial \rho/\partial P)_T^{-1}$) are determined for $\phi>0.58$ by the equalizing of potential interaction of contacting molecules and their thermal energy. Consequently, the universal line of the inverse reduced volume fluctuations splits into a bundle of curves specific for each isotherm as it is visible in Fig.~\ref{fgr:numeth}.

This line of reasoning can be also supported from the point of view of a general approach of statistical physics to liquid structure. First of all, the considered parameter is directly connected with the compressibility equation \cite{Hansen} represented in the form
\begin{equation}
\nu^{-1}=
1+\rho\frac{N_A}{\mu_0}\int\limits_{0}^{\infty}\left[g(r)-1\right]2\pi r^2dr,
\label{compeq}
\end{equation}
where the term multiplied by the density, is so-called the Kirkwood-Buff integral $G(\rho,T)$ taken over the pair correlation function $h(r)=g(r)-1$ with the radial distribution function $g(r)$. $N_A$ is Avogadro's constant. As it has been analysed in detail by Ben-Naim \cite{BenNaim2008}, the value of the Kirkwood-Buff integral for a one-component system has a strong dependence on the macroscopic density but 
loses the microscopic information on the details of intermolecular interactions. 

Introducing the dimensionless length\cite{BenNaim2008} $r'=r/\sigma$, where $\sigma$ is the van der Waals diameter and the van der Waals volume\cite{Bondi1964} $V_{vdW}=\pi\sigma^3/6$, we can reduce the compressibility equation along an isotherm (i.e. for a fixed $T$, which is therefore omitted in the further notation) to the form 
$$
\nu^{-1}=1+\phi\frac{6}{\pi}G(\rho).
$$
Here the Kirkwood-Buff integral $G(\rho)$ has the same scale as in\cite{BenNaim2008}, where it has been calculated for model systems (hard spheres and the dense Lennard-Jones liquid) directly from the first principles of statistical physics.  

The estimation of $G(\rho)$ for the considered case of dibromomethane around the revealed transition ($\log(\nu)\approx0.39$, $\phi\approx0.574$) gives $G=-0.89$. Although the comparison is definitely semi-qualitative (due to a certain difference between the model and the real liquids), this value is within the boundary of the discussed range\cite{BenNaim2008} corresponding to the universal asymptotic convergence of integral's value, which originates from a packing approaching to the close one.

\begin{figure}[t]
\centering
  \includegraphics[width=\columnwidth]{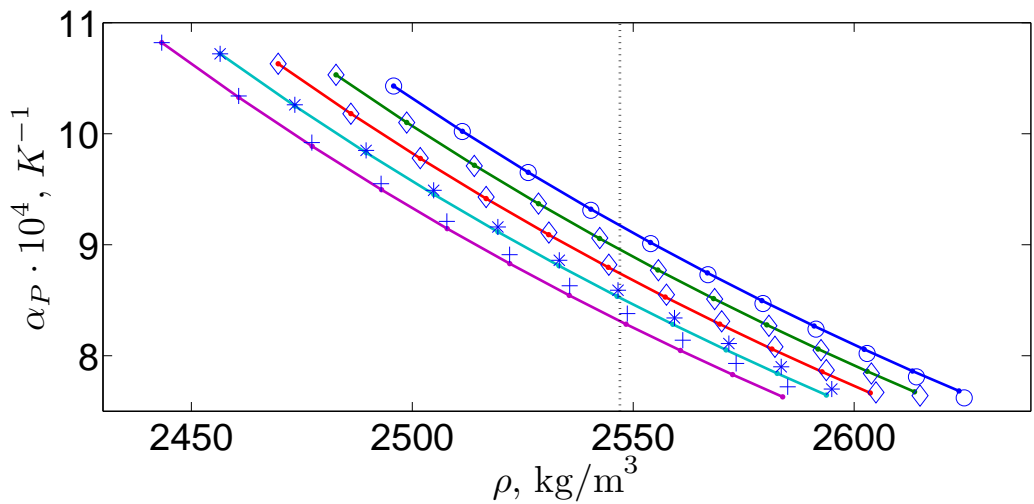}
  \caption{The measured (markers are the same as in Fig.~\ref{fgr:numeth}) and predicted values of the isobaric thermal expansion coefficient for dibromomethane. The vertical dotted line corresponds to the same characteristic density as in Fig.~\ref{fgr:allnu}.
	\label{fgr:alphameth}}
\end{figure}

The comparison of densities corresponding to the transition from the temperature-independent exponential to the temperature-dependent power-law fluctuations and the densities at the pressures corresponding to the intervals of $\left(\partial \alpha_p/\partial T\right)_P=0$ location (Fig.~\ref{fgr:intersect}) demonstrates their practical coincidence.

Thus, we can conclude that the crossing of expansivity isotherms is directly connected with the described structural transition of a normal liquid with holes (loose packing) to the liquid with irregularly closely packed particles and the line of crossing demarcate these two states. At the same time, it is not a ``phase transition'' in the strict sense due to the absence of a discontinuity of  thermodynamic functions. Instead, it is more related to the hypothesis discussed by Randzio\cite{Randzio1986} about different excitability of the inner degrees of freedom and various lengths of the minimal separation between particles with different energies/temperatures as a key factor of different anomalies of the isobaric expansion coefficient.

Finally, it should be pointed out that the deviations from the exponential behaviour are small for the considered substances and Eq. (\ref{alphc}) provides a reasonable accuracy for the values of expansivity even after the crossing region.
 In fact, the relative error does not exceed $4\%$ for 100~MPa for dibromomethane, see Fig.~\ref{fgr:alphameth}. Here Eqs.~(\ref{nuexp})--(\ref{alphc}) are used with the constants $\kappa=0.0035~\mathrm{m^3/kg}$, $b=-4.94$, the referent sets $\rho_0$, $\alpha_P^0$ are taken at the normal pressure from \cite{Chorazewski2014-2}.

The accuracy is even better for longer chained dibromoalkanes: within the range of $1.5\%$, see the recently published work\cite{Chorazewski2015}. This effect can be explained by the decreasing of the proximity effect for substances, which contain lower proportion of halogen atoms in their molecular composition. 

\section{Conclusion and outlook}

The knowledge of thermal expansion coefficient is crucial for the designing of chemical processes as well as for the progress of thermodynamic theories. Herein, the new isothermal equation of state (FT-EoS) provides the method of its calculation and allows the discussion of its behaviour in connection with the fluctuational and structural properties of a liquid. 

It is easy to show that the first term in Eq.~\ref{alphc}  can be rewritten in the form $\alpha_p^0\left[\rho_0\nu(\rho_0)\left/\rho\nu(\rho)\right.\right]$
i.e. as a ratio of combinations $\rho\nu(\rho)$, which were first considered in the work \cite{Huang1987} as the ``reciprocal compressibility''. It has been shown there that this quantity itself may exhibit isotherm crossing due to a difference of structural changes (second derivative of the entropy) and force changes (second derivative of the internal energy)  under the elevated pressure. This observation motivated the analysis of the isotherm crossing phenomenon in connection with the characteristic Zeno line and the Boyle volume \cite{Boushehri1993}. Recently, the attention to this topic is revitalized \cite{Boshkova2010,Nedostup2015} as a background for the revealing configurational and thermal components in the statistical physics description of liquid state.

In addition, the approach considered  has a certain interplay with the problems of transitions between quantitatively different diffusion regimes in soft matter systems at growing density. As it has been discussed in \cite{Goncharov2013}, the inverse reduced fluctuations $\nu$ correspond to the reduced coefficient of self-diffusion. At the same time, recent numerical studies \cite{Ghosh2015,Shin2015,Schnyder2015} show the transition from the normal (with exponentially decaying correlations) diffusion to the suppressed one governed by the volume occupancy. Due to the similarity of description and observed effects, one can suppose that the sophisticated methods developed in the theory of anomalous stochastic processes may improve understanding of classical liquid state physics and, {\it visa versa}, the approach based on the inverse reduced fluctuations, which are determined by the macroscopic thermodynamic values (the density, the isothermal compressibility) may simplify the study of complex 
diffusion processes replacing much more difficult tracing of individual walkers in the case of real substances.

Finally, FT-EoS has also several practical advantages over the other methods aimed on the modelling of the thermal expansivity in the case of liquids. First, the proposed model is accurate and reliable over the liquid region. Moreover, its structure is also relatively simple, which yields reliable thermal expansivity values by fitting two parameters only, and does not require measurements at elevated pressures. Thus, our new model also satisfies the demands on accuracy for chemical engineering applications.

%\bibliographystyle{elsarticle-num}
%\bibliography{bibalpha} 

\begin{thebibliography}{10}
\expandafter\ifx\csname url\endcsname\relax
  \def\url#1{\texttt{#1}}\fi
\expandafter\ifx\csname urlprefix\endcsname\relax\def\urlprefix{URL }\fi
\expandafter\ifx\csname href\endcsname\relax
  \def\href#1#2{#2} \def\path#1{#1}\fi

\bibitem{Ho}
C.~Y. Ho, R.~E. Taylor (Eds.), Thermal Expansion of Solids, Cindas Data Series
  on Material Properties, V. I-4., ASM International, 1998.

\bibitem{Bridgman1913}
P.~W. Bridgman, Thermodynamic properties of twelve liquids between 20$^o$ and
  80$^o$ and up to 12000 kgm. per sq. cm., P. Am. Acad. Arts Sci. 49 (1913)
  3--114.
\newblock \href {http://dx.doi.org/10.2307/20025445}
  {\path{doi:10.2307/20025445}}.

\bibitem{Randzio1994}
S.~L. Randzio, J.-P.~E. Grolier, J.~R. Quint, D.~J. Eatough, E.~A. Lewis, L.~D.
  Hansen, n-hexane as a model for compressed simple liquids, Int. J.
  Thermophys. 15 (1994) 415--441.
\newblock \href {http://dx.doi.org/10.1007/BF01563706}
  {\path{doi:10.1007/BF01563706}}.

\bibitem{Troncoso2009}
J.~Troncoso, C.~A. Cerdeirina, P.~Navia, Y.~A. Sanmamed,
  D.~Gonz{\'a}lez-Salgado, L.~Roman{\'\i}, Unusual behavior of the
  thermodynamic response functions of ionic liquids, J. Phys. Chem. Lett. 1~(1)
  (2009) 211--214.
\newblock \href {http://dx.doi.org/10.1021/jz900049g}
  {\path{doi:10.1021/jz900049g}}.

\bibitem{Taravillo2003}
M.~Taravillo, V.~G. Baonza, M.~C{\'a}ceres, J.~Nunez, Thermodynamic
  regularities in compressed liquids: I. the thermal expansion coefficient, J.
  Phys.: Condensed Matter 15 (2003) 2979.
\newblock \href {http://dx.doi.org/10.1088/0953-8984/15/19/302}
  {\path{doi:10.1088/0953-8984/15/19/302}}.

\bibitem{Tellez2011}
P.~T{\'e}llez-Arredondo, M.~Medeiros, M.~M. Pi{\~n}eiro, C.~A. Cerdeiri{\~n}a,
  Loci of extrema of thermodynamic response functions for the Lennard--Jones
  fluid, Molecular Physics 109 (2011) 2443--2449.
\newblock \href {http://dx.doi.org/10.1080/00268976.2011.619505}
  {\path{doi:10.1080/00268976.2011.619505}}.

\bibitem{Chorazewski2010}
M.~Chor{\c a}\.zewski, J.-P.~E. Grolier, S.~L. Randzio, Isobaric thermal
  expansivities of toluene measured by scanning transitiometry at temperatures
  from (243 to 423) K and pressures up to 200 MPa, J. Chem. Eng. Data 55 (2010)
  5489--5496.
\newblock \href {http://dx.doi.org/10.1021/je100657n}
  {\path{doi:10.1021/je100657n}}.

\bibitem{Troncoso2011}
J.~Troncoso, P.~Navia, L.~Roman{\'\i}, D.~Bessieres, T.~Lafitte, On the
  isobaric thermal expansivity of liquids, J. Chem. Phys. 134 (2011) 094502.
\newblock \href {http://dx.doi.org/10.1063/1.3549828}
  {\path{doi:10.1063/1.3549828}}.

\bibitem{Randzio2014}
S.~L. Randzio, J.-P.~E. Grolier, M.~Chor{\c a}\.zewski, High-pressure Maxwell
  relations measurements, in: E.~Wilhelm, T.~Letcher (Eds.), Volume
  Properties: Liquids, Solutions and Vapours, Royal Society of Chemistry, 2014,
  pp. 414--438.

\bibitem{Randzio1995}
S.~L. Randzio, U.~K. Deiters, Thermodynamic testing of equations of state of
  dense simple liquids, Ber. Bunsenges. Phys. Chem. 99 (1995) 1179--1186.
\newblock \href {http://dx.doi.org/10.1002/bbpc.199500057}
  {\path{doi:10.1002/bbpc.199500057}}.

\bibitem{Deiters1995}
U.~K. Deiters, S.~L. Randzio, The equation of state for molecules with shifted
  Lennard-Jones pair potentials, Fluid Phase Equilib. 103 (1995) 199--212.
\newblock \href {http://dx.doi.org/10.1016/0378-3812(94)02577-N}
  {\path{doi:10.1016/0378-3812(94)02577-N}}.

\bibitem{Chauhan2013}
R.~S. Chauhan, P.~Singh, C.~P. Singh, Analysis of thermoelastic properties and
  equation of state for cyclopentane, tetramethylsilane and 2,
  3-dimethylbutane, Phys. Chem. Liq. 51 (2013) 294--301.
\newblock \href {http://dx.doi.org/10.1080/00319104.2012.713551}
  {\path{doi:10.1080/00319104.2012.713551}}.

\bibitem{Alba1985}
C.~Alba, L.~Ter~Minassian, A.~Denis, A.~Soulard, Reduction into a rational
  fraction of a thermodynamic property of the liquid state: Experimental
  determinations in the case of $CO_2$ and $n$-butane. extension to the other
  properties, J. Chem. Phys. 82 (1985) 384--393.
\newblock \href {http://dx.doi.org/10.1063/1.448757}
  {\path{doi:10.1063/1.448757}}.

\bibitem{Baonza1993}
V.~G. Baonza, M.~Caceres, J.~Nunez, Extended analytical equation of state for
  liquids from expansivity data analysis, J. Phys. Chem. 97 (1993)
  10813--10817.
\newblock \href {http://dx.doi.org/10.1021/j100143a047}
  {\path{doi:10.1021/j100143a047}}.

\bibitem{Goncharov2013}
A.~L. Goncharov, E.~B. Postnikov, V.~V. Melent'ev, Limits of structure
  stability of simple liquids revealed by study of relative fluctuations, Eur.
  Phys. J. B 86 (2013) 357.
\newblock \href {http://dx.doi.org/10.1140/epjb/e2013-40111-7}
  {\path{doi:10.1140/epjb/e2013-40111-7}}.

\bibitem{Postnikov2014}
E.~B. Postnikov, A.~L. Goncharov, V.~V. Melent'ev, Tait equation revisited from
  the entropic and fluctuational points of view, Int. J. Thermophys. 35 (2014)
  2115--2123.
\newblock \href {http://dx.doi.org/10.1007/s10765-014-1747-5}
  {\path{doi:10.1007/s10765-014-1747-5}}.

\bibitem{Chorazewski2015}
M.~Chor{\c{a}}{\.z}ewski, E.~B. Postnikov, Thermal properties of compressed
  liquids: Experimental determination via an indirect acoustic technique and
  modeling using the volume fluctuations approach, Int. J. of Therm. Sci. 90
  (2015) 62--69.
\newblock \href {http://dx.doi.org/10.1016/j.ijthermalsci.2014.11.028}
  {\path{doi:10.1016/j.ijthermalsci.2014.11.028}}.

\bibitem{Chorazewski2014-2}
M.~Chor{\c{a}}{\.z}ewski, J.~Troncoso, J.~Jacquemin, Thermodynamic properties
  of dichloromethane, bromochloromethane and dibromomethane under elevated
  pressure: Experimental results and SAFT-VR Mie predictions, Industrial \&
  Engineering Chemistry Research 54 (2014) 720--730.
\newblock \href {http://dx.doi.org/10.1021/ie5038903}
  {\path{doi:10.1021/ie5038903}}.

\bibitem{Chorazewski2015-2}
E.~B. Chor{\c{a}}{\.z}ewski, M.and~Postnikov, K.~Oster, I.~Polishuk,
  Thermodynamic properties of 1,2-dichloroethane and 1,2-dibromoethane under
  elevated pressures: Experimental results and predictions of a novel
  DIPPR-based version of FT-EoS, PC-SAFT and CP-PC-SAFT, Ind. Eng. Chem. Res.
  54 (2015) 9645--9656.
\newblock \href {http://dx.doi.org/10.1021/acs.iecr.5b02626}
  {\path{doi:10.1021/acs.iecr.5b02626}}.

\bibitem{Kehiaian1983}
H.~V. Kehiaian, Group contribution methods for liquid mixtures: a critical
  review, Fluid Phase Equilib. 13 (1983) 243--252.
\newblock \href {http://dx.doi.org/10.1016/0378-3812(83)80098-3}
  {\path{doi:10.1016/0378-3812(83)80098-3}}.

\bibitem{Onoda1990}
G.~Y. Onoda, E.~G. Liniger, Random loose packings of uniform spheres and the
  dilatancy onset, Phys. Rev. Lett. 64 (1990) 2727--2730.
\newblock \href {http://dx.doi.org/10.1103/PhysRevLett.64.2727}
  {\path{doi:10.1103/PhysRevLett.64.2727}}.

\bibitem{Aste2008}
T.~Aste, T.~Di~Matteo, Structural transitions in granular packs: statistical
  mechanics and statistical geometry investigations, Eur. Phys. J. B 64~(3)
  (2008) 511--517.
\newblock \href {http://dx.doi.org/10.1140/epjb/e2008-00224-8}
  {\path{doi:10.1140/epjb/e2008-00224-8}}.

\bibitem{Heyes2006}
D.~M. Heyes, H.~Okumura, Equation of state and structural properties of the
  Weeks-Chandler-Andersen fluid, J. Chem. Phys. 124 (2006) 164507.
\newblock \href {http://dx.doi.org/10.1063/1.2176675}
  {\path{doi:10.1063/1.2176675}}.

\bibitem{Hansen}
J.~P. Hansen, I.~R. McDonald, Theory of Simple Liquids, London: Academic Press,
  2006.

\bibitem{BenNaim2008}
A.~Ben-Naim, The Kirkwood--Buff integrals for one-component liquids, J. Chem.
  Phys. 128 (2008) 234501.
\newblock \href {http://dx.doi.org/10.1063/1.2938859}
  {\path{doi:10.1063/1.2938859}}.

\bibitem{Bondi1964}
A.~Bondi, van der Waals volumes and radii, J. Phys. Chem. 68 (1964) 441--451.
\newblock \href {http://dx.doi.org/10.1021/j100785a001}
  {\path{doi:10.1021/j100785a001}}.

\bibitem{Randzio1986}
S.~L. Randzio, An attempt to explain thermal properties of liquids at high
  pressures, Phys. Lett. A 117~(9) (1986) 473--476.
\newblock \href {http://dx.doi.org/10.1016/0375-9601(86)90406-8}
  {\path{doi:10.1016/0375-9601(86)90406-8}}.

\bibitem{Huang1987}
Y.-H. Huang, J.~P. O'Connell, Corresponding states correlation for the
  volumetric properties of compressed liquids and liquid mixtures, Fluid Phase
  Equilib. 37 (1987) 75--84.
\newblock \href {http://dx.doi.org/10.1016/0378-3812(87)80044-4}
  {\path{doi:10.1016/0378-3812(87)80044-4}}.

\bibitem{Boushehri1993}
A.~Boushehri, F.~M. Tao, E.~A. Mason, Common bulk modulus point for compressed
  liquids, J. Phys. Chem. 97 (1993) 2711--2714.
\newblock \href {http://dx.doi.org/10.1021/j100113a037}
  {\path{doi:10.1021/j100113a037}}.

\bibitem{Boshkova2010}
O.~L. Boshkova, U.~K. Deiters, Soft repulsion and the behavior of equations of
  state at high pressures, Intern. J. Thermoph. 31 (2010) 227--252.
\newblock \href {http://dx.doi.org/10.1007/s10765-010-0727-7}
  {\path{doi:10.1007/s10765-010-0727-7}}.

\bibitem{Nedostup2015}
V.~I. Nedostup, Classical ideal curves in the phase diagrams for simple
  substances, High Temper. 53 (2015) 62--67.
\newblock \href {http://dx.doi.org/10.1134/S0018151X14060091}
  {\path{doi:10.1134/S0018151X14060091}}.

\bibitem{Ghosh2015}
S.~K. Ghosh, A.~G. Cherstvy, R.~Metzler, Non-universal tracer diffusion in
  crowded media of non-inert obstacles, Phys. Chem. Chem. Phys. 17 (2015)
  1847--1858.
\newblock \href {http://dx.doi.org/10.1039/C4CP03599B}
  {\path{doi:10.1039/C4CP03599B}}.

\bibitem{Shin2015}
J.~Shin, A.~G. Cherstvy, R.~Metzler, Self-subdiffusion in solutions of
  star-shaped crowders: non-monotonic effects of inter-particle interactions, New J. Phys. 17 (2015) 113028
\newblock \href {http://dx.doi.org/10.1088/1367-2630/17/11/113028}
  {\path{doi:10.1088/1367-2630/17/11/113028}}.

\bibitem{Schnyder2015}
S.~K. Schnyder, M.~Spanner, F.~H{\"o}fling, T.~Franosch, J.~Horbach, Rounding
  of the localization transition in model porous media, Soft Matter 11 (2015)
  701--711.
\newblock \href {http://dx.doi.org/10.1039/C4SM02334J}
  {\path{doi:10.1039/C4SM02334J}}.

\end{thebibliography}

\end{document}